\documentclass[a4paper,12pt]{article}
\usepackage{graphicx}
\title{The Y(4260) as a $J/\psi K \bar{K}$ system.}
\author{A. Mart\'inez Torres$^1$\footnote{amartine@ific.uv.es},
K. P. Khemchandani$^2$\footnote{kanchan@teor.fis.uc.pt}, D. Gamermann$^1$\footnote{gamerman@ific.uv.es},\\ and E.~Oset$^1$\footnote{oset@ific.uv.es} \\
{\small{\it $^1$Departamento de F\'{\i}sica Te\'orica and IFIC,
Centro Mixto Universidad de Valencia-CSIC,}}\\
{\small{\it Institutos de
Investigaci\'on de Paterna, Aptdo. 22085, 46071 Valencia, Spain}}\\
{\small{\it $^2$ Centro de F\'isica Te\'orica, Departamento de F\'isica,}}\\
{\small{\it Universidade de Coimbra, P-3004-516 Coimbra, Portugal. 
}}\\
}
\date{\today}

\newcommand{\dmu}{\partial_\mu}

\newcommand{\lc}{{\cal L}}

\newcommand{\sqd}{\sqrt{2}}

\newcommand{\be}{\begin{eqnarray}}
\newcommand{\ee}{\end{eqnarray}}
\newcommand{\nn}{\nonumber}

\begin{document}

\maketitle

\begin{abstract}
A study of the $J/\psi \pi \pi$ and $J/\psi K \bar{K}$ systems, treating them as coupled channels, has been made by solving the Faddeev equations,  with the purpose of investigating the possibility of generation of the $J^{PC} = 1^{--}$, $Y(4260)$ resonance due to the interaction between these three mesons. In order to do this, we start by solving the Bethe-Salpeter equation for the two pseudoscalar and for the vector-pseudocalar meson systems using the amplitudes obtained from the lowest order chiral Lagrangians as potentials. With the $t$-matrices generated from these potentials, which contain the poles of the $\sigma$, $f_{0}(980)$ and $a_{0}(980)$ resonances for the pseudoscalar-pseudoscalar system and the pole of the $X(3872)$, alongwith other new charmed resonant states, for the vector-pseudoscalar system, we solve the Faddeev equations. As a result, we get a peak around 4150 MeV with a width  $\sim$ 90 MeV when the invariant mass of the two pseudoscalars is close to that of the $f_0 (980)$. 
\end{abstract}

\section{Introduction}
An enhancement in the data for the $\pi^+\pi^- J/\psi$ invariant mass spectrum was found near 4.26 GeV by the BABAR collaboration in a study of the $e^+ e^- \rightarrow \gamma_{ISR} \pi^+\pi^- J/\psi$ process\cite{babar1}. A fit to this data set was made by assuming a resonance with 4.26 GeV of mass and 50 to 90 MeV of width \cite{babar1}. The resonance was named as the $Y(4260)$ and it was found to be of $J^{PC} =1^{--}$ nature. Later on, an accumulation of events with similar characteristics in  the $\pi^+\pi^- J/\psi$, $\pi^0 \pi^0 J/\psi$ and the $K^+K^- J/\psi$ mass spectra was reported by the CLEO collaboration \cite{cleo1,cleo2}, thus confirming the results from BABAR. Following these works, the BELLE collaboration obtained the cross sections for the 
$e^+ e^- \rightarrow  \pi^+\pi^- J/\psi$ reaction in the 3.8 to 5.5 GeV region \cite{belle1}, by keeping all the interactions in the final state in $S$-wave, and found a peak at 4.26 GeV and a bump around 4.05 GeV.

Although the $Y(4260)$ does not seem to fit in to the charmonium spectrum  of the particle data group \cite{pdg} known up to $\sim$ 4.4 GeV, a proposal to accommodate it as a $4s$ state has been made in \cite{felipe}. Several other suggestions have been made for the interpretation of this state, for example, the authors of \cite{tetraquark} propose it to be a tetra-quark state, others propose a hadronic molecule of $D_{1} D$, $D_{0} D^*$ \cite{Ding,Albuquerque}, $\chi_{c1} \omega$ \cite{Yuan}, $\chi_{c1} \rho$ \cite{liu} and yet another idea is that it is a hybrid charmonium \cite{zhu} or charm baryonium \cite{Qiao}, etc. Within the available experimental information, none of these suggestions can be completely ruled out and its not clear if $Y(4260)$ possesses any of these structures dominantly or is a mixture of all of them. In Refs. \cite{eef1,eef2,eef3} the authors call the attention of the readers to the presence of the opening of the $D^*_s \bar{D}^*_s$ channel very close to the peak position of the $Y(4260)$ in the updated data from BABAR
\cite{babarupdate} and associate the peak corresponding to $Y(4260)$ to a $D^*_s \bar{D}^*_s$ cusp. A fit to the data from \cite{babar1,babarupdate} has been made in \cite{eef2} and additional presence of a rather broad bump around 4.35 GeV has been proposed.

There are some peculiarities in the experimental findings which motivate us to carry out a study of the $J/\psi \pi \pi$ system. There is no enhancement found around 4.26 GeV in the process with the $D^* \bar{D}^*$ \cite{no4260} or other hadron final states \cite{pdg1} and it is concluded that $Y(4260)$ has an unusually strong coupling to the $\pi \pi J/\psi$ final state \cite{babar1,cleo1,cleo2,belle1}. Further, the data on the invariant mass of the $\pi \pi$ subsystem obtained by the Belle collaboration  \cite{belle1}, for total energy range, 3.8-4.2 GeV, 4.2-4.4 GeV and 4.4-4.6 GeV, have curious features. The $\pi \pi$ mass distribution data in 3.8-4.2 GeV and 4.4-4.6 GeV seem to follow the phase space, however, that corresponding to the 4.2-4.4 GeV total energy differs significantly from the phase space and shows an enhancement near $m_{\pi \pi} =$ 1 GeV. Do these findings indicate that the $Y(4260)$ has a strong coupling to $f_0(980) J/\psi$, similar to the $X(2175)$ to the $\phi f_{0}(980)$ \cite{babar2,bes}? It is interesting to recall that the $X(2175)$ was found as a dynamically generated resonance in the $\phi K \bar{K}$ system \cite{us2175,alvarez} with the $K \bar{K}$ subsystem possessing the characteristics of the $f_0(980)$. Similarly, the $Y(4660)$ \cite{belle2} has been suggested as a $\psi^\prime f_0(980)$ resonance \cite{guo}. In order to find an answer to this question, we  have solved the Faddeev equations for the $J/\psi \pi \pi$  and $J/\psi K \bar{K}$ coupled channels and we discuss the formalism and results of our study in the following sections.

\section{Formalism}
\begin{figure}[h!]
\begin{center}
\includegraphics[width=8cm]{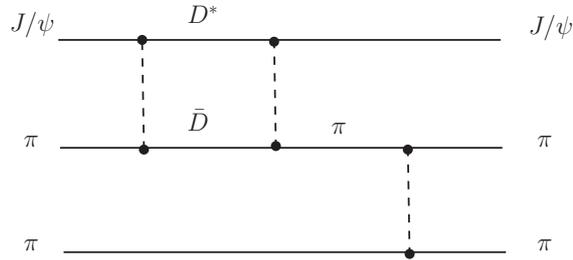}
\caption{A three-body interaction diagram where the $J/\psi \pi$ interaction proceeds through $\bar{D} D^*$ coupled channel.}
\label{fig1}
\end{center}
\end{figure}

\begin{figure}[h!]
\begin{center}
\includegraphics[width=12cm]{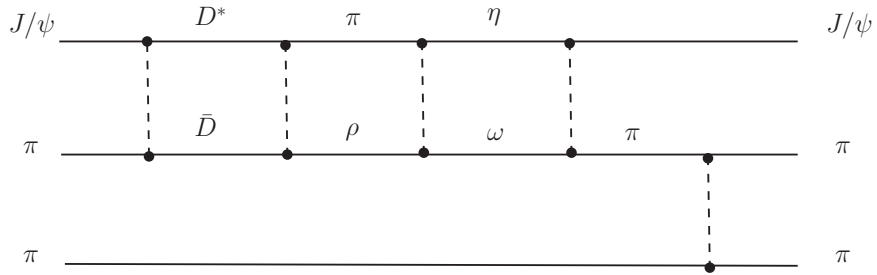}
\caption{Another possible contribution of the $J/\psi \pi$ amplitude, through loops of other coupled channels, to the three-body interaction.}
\label{fig2}
\end{center}
\end{figure}

In our earlier study of the $\phi K \bar{K}$ system we found the dynamical generation of the $X(2175)$ resonance \cite{us2175}. The study was carried out by solving the Faddeev equations for the three-meson system using chiral Lagrangians for interaction of the constituent mesons.
There are some similarities between the $X(2175)$ and the $Y(4260)$. Both resonances are of $J^{PC}=1^{--}$ nature. The $X(2175)$ was found in the $\phi f_0(980)$ cross sections \cite{babar2,bes} and a study of this system using chiral dynamics required calculations for the  $\phi K \bar{K}$ system since the $f_0(980)$ is basically a $K \bar{K}$ molecule in such a formalism. 
The $Y(4260)$ has been found in a system of a vector and two pseudoscalar mesons, with the two pseudoscalars interacting in the $S$-wave and with their invariant mass showing a dominant peak around 1 GeV in the $Y(4260)$ region. This hints towards a possibility of clustering of the two pions to form the $f_0(980)$.
If the two pions rearranged themselves to form the $f_0(980)$ resonance, the  $Y(4260)$ would be about 200 MeV above the $J/\psi f_0$ threshold just as in case of the $X(2175)$ which is about 200 MeV above the $\phi f_0(980)$ threshold. 
Besides, the diagonal term of the potential obtained from chiral Lagrangian for $J/\psi \pi$ is zero just as the one for the $\phi K$ (or $\phi \pi$) interaction. However, the $\phi \pi$ (or $\phi K$) scattering matrix  is non-zero due to loops of the non-diagonal (coupled channel) $\phi \pi \rightarrow \bar{K} K^* (K \bar{K}^*)$  terms. Similarly, the  $J/\psi \pi \rightarrow J/\psi \pi$ at the  lowest order is null but the scattering matrix is formed through iterations of the potential involving non diagonal transitions within the coupled channels, like $J/\psi \pi\to \bar{D} D^*\to J/\psi \pi$. This would give rise to three-body diagrams of the kind shown in Figs. \ref{fig1} and \ref{fig2}. All these mentioned similarities between the $X(2175)$ and the $Y(4260)$, and the experimental findings of $Y(4260)$ with seemingly strong coupling to the $J/\psi \pi \pi$ channel motivate us to carry out  a three-body calculation of the $J/\psi \pi \pi$ system.

We study the $J/\psi \pi \pi$ and $J/\psi K \bar{K}$ systems as coupled channels in the isospin 0 base and by considering all the interactions in  $S$-wave. For this we solve the Faddeev equations in the formalism developed and used earlier to study three-meson system and two meson one baryon systems \cite{us2175,us_pkbarn1,us_pkbarn2,us_ppn1,us_ppn2,us_pkn}. The different two-meson one baryon systems which we have studied so far are the $\pi \bar{K} N$, $\pi \pi N$ and the $\pi K N$ system, including the corresponding coupled channels in each case. In case of $S = -1$, i.e., $\pi \bar{K} N$ system, we find dynamical generation of four isospin 1 and two isospin 0 resonances \cite{us_pkbarn1}, which we relate to the $\Sigma(1560)$, $\Sigma(1620)$,$\Sigma(1660)$,$\Sigma(1770)$, $\Lambda(1600)$ and $\Lambda(1810)$ of the particle data group ($PDG$) \cite{pdg}. In $S=0$ case, we find evidence of strong coupling of $N^*(1710)$,  $N^*(2100)$, $\Delta(1910)$ to two meson one baryon channels \cite{us_ppn1,us_ppn2} and also find another isospin 1/2 baryon resonance with spin-parity = $1/2^+$ around 1920 MeV \cite{us_ppn2}. The latter one was earlier found in a non-relativistic study of the $K \bar{K} N$ system using a variational method  \cite{jido} and a signature of this resonance actually seems to be present in experimental data on $\gamma p \rightarrow K^+ \Lambda$ (see Ref. \cite{eulogio}). We also checked the possibility of generation of $S = 1$ resonances from three-body dynamics and found a broad bump around 1700 MeV \cite{us_pkn}.

The formalism consists of calculating the equation
\begin{equation}
T_R = T^{\,12}_R + T^{\,13}_R + T^{\,21}_R + T^{\,23}_R + T^{\,31}_R + T^{\,32}_R,\label{full}
\end{equation}
where,
\begin{eqnarray} \nonumber
T^{\,12}_R&=&t^1g^{12}t^2+t^1\Big[G^{\,121\,}T^{\,21}_R+G^{\,123\,}T^{\,23}_R\Big] \nonumber\\
T^{\,13}_R&=&t^1g^{13}t^3+t^1\Big[G^{\,131\,}T^{\,31}_R+G^{\,132\,}T^{\,32}_R\Big] \nonumber\\
T^{\,21}_R&=&t^2g^{21}t^1+t^2\Big[G^{\,212\,}T^{\,12}_R+G^{\,213\,}T^{\,13}_R\Big] \label{Trest}\\
T^{\,23}_R&=&t^2g^{23}t^3+t^2\Big[G^{\,231\,}T^{\,31}_R+G^{\,232\,}T^{\,32}_R\Big] \nonumber\\
T^{\,31}_R&=&t^3g^{31}t^1+t^3\Big[G^{\,312\,}T^{\,12}_R+G^{\,313\,}T^{\,13}_R\Big] \nonumber\\
T^{\,32}_R&=&t^3g^{32}t^2+t^3\Big[G^{\,321\,}T^{\,21}_R+G^{\,323\,}T^{\,23}_R\Big],\nonumber  
\end{eqnarray}
which can be related to the Faddeev partitions as
\begin{eqnarray} \nonumber
T^{\,1}&=&t^1 \delta^3(\vec{k}^{\,\prime}_1-\vec{k}_1) + T^{\,12}_R + T^{\,13}_R \\ \nonumber
T^{\,2}&=&t^2 \delta^3(\vec{k}^{\,\prime}_2-\vec{k}_2) + T^{\,21}_R + T^{\,23}_R  \\ 
T^{\,3}&=&t^3 \delta^3(\vec{k}^{\,\prime}_3-\vec{k}_3) + T^{\,31}_R + T^{\,32}_R.\label{fe}
\end{eqnarray}
In Eqs. (\ref{fe}), $\vec{k}^{\,\prime}_i (\vec{k}_i)$ is the momentum of the $ith$ particle in the final (initial) global center of mass system. The $t^i \delta^3(\vec{k}^{\,\prime}_i-\vec{k}_i)$ terms in Eqs. (\ref{fe}) correspond to  three-body diagrams with an interaction $t^i$ which represents a two body $t$-matrix, $t^i = v^i + v^i g^{i \,\prime} t^i$, between the $jth$ and $kth$ particles, with $j \ne k \ne i$, and with the $ith$ particle being a spectator. Thus, the superindex on the $t^i$'s indicates the non-interacting particle. Such diagrams correspond to  disconnected three-body diagrams, and removing them one is left with all the connected diagrams of the Faddeev equations, the sum of which we  denote as $T_R$ (where the subscript $R$ denotes the ``rest'' of the diagrams). Hence the $T^{\,ij}_R$ matrices correspond to the sum of all the connected diagrams with the last two  interactions described by $t^i$ and $t^j$.

These $t^l$-matrices
in our formalism are obtained by solving coupled channel Bethe-Salpeter equations with the potentials obtained from chiral Lagrangians. These potentials can be written as a sum of two terms, one depending only on the center of mass energy of the interacting pair and other depending on  off-shell variables. As has been discussed elaborately in our previous works \cite{us2175,us_pkbarn1,us_pkbarn2,us_ppn1,us_ppn2,us_pkn}, interestingly, we find the contribution of these off-shell parts of the potentials, and hence of the $t$-matrices, to cancel exactly the three-body forces originating from the same chiral Lagrangian in the SU(3) limit. In a realistic case, we found that the contributions from the off-shell part of the $t^l$-matrices  together with these contact three-body terms were negligibly small. An explicit analytic proof of these cancellations has been shown for the case of two meson-one baryon systems in \cite{us_ppn1} and for the case of three mesons in \cite{us2175}. This important finding allows us to solve the Faddeev equations by using the on-shell part of the two-body $t^l$-matrices, which are independent of the unphysical off-shell parts. Thus, we calculate Eqs. (\ref{Trest}) with the $t^l$-matrices depending on the invariant mass of the interacting pair.

Eqs. (\ref{Trest}) are solved as a function of the total energy, $\sqrt{s}$ and the invariant mass of the $23$ system, $\sqrt{s_{23}}$. In our case, labeling $J/\psi$ as the particle $1$ and $K \bar{K}$ (and $\pi\pi$) as the particles $2$ and $3$, $\sqrt{s_{23}}$ is the invariant mass of the two pseudoscalars. We define the momenta and other invariant masses in terms of $\sqrt{s}$ and $\sqrt{s_{23}}$ 
as shown in detail in \cite{us_ppn1}.

In Eqs. (\ref{Trest}), the first term, $t^i g^{ij}t^j$ (with $i \neq j$) represents the simplest possible three-body connected diagram which contains two $t$-matrices, where the $g^{ij}$ matrix elements are three-body Green's functions, defined as
\begin{equation}
g^{ij} (\vec{k}^\prime_i, \vec{k}_j)=\Bigg(\frac{1}{2E_k(\vec{k}^\prime_i+\vec{k}_j)}\Bigg)\frac{1}{\sqrt{s}-E_i
(\vec{k}^\prime_i)-E_j(\vec{k}_j)-E_k(\vec{k}^\prime_i+\vec{k}_j)+i\epsilon},
\end{equation}
with $E_r = \sqrt{\vec{k_r}^2 + m_r^2}$ and $m_r$ is the mass of the $rth$ particle of that coupled channel to which the element of the matrix corresponds.
The mathematical expression for the next order diagram, that is the one with three $t$-matrices, is written as 
$t^i G^{ijk} t^j g^{jk} t^k$, where the $G^{ijk}$ matrix is 
\begin{equation}
G^{i\,j\,k}  =\int\frac{d^3 k^{\prime\prime}}{(2\pi)^3}\tilde{g}^{ij} \cdot F^{i\,j \,k}\hspace{1cm}{(i \ne j, j \ne k =1,2,3)}
\end{equation}
with the elements of the $\tilde{g}^{ij}$ being 
\begin{eqnarray}
\tilde{g}^{ij} (\vec{k}^{\prime \prime}, s_{lm}) = \frac{1}
{2E_l(\vec{k}^{\prime\prime})} \frac{1}{2E_m(\vec{k}^{\prime\prime})} \frac{1}{\sqrt{s_{lm}}-E_l(\vec{k}^{\prime\prime})-E_m(\vec{k}^{\prime\prime})
+i\epsilon} \label{eq:G} \\
\hspace{2.3cm}(i \ne l \ne m, i \ne j =1,2,3) \nonumber
\end{eqnarray}
and the matrix $F^{i\,j\,k}$, with explicit variable dependence, is written as 
\begin{equation}
F^{i\,j\,k} (\vec{k}^{\prime \prime},\vec{k}^\prime_j, \vec{k}_k,  s^{k^{\prime\prime}}_{ru})= t^{j}(s^{k^{\prime\prime}}_{ru}) g^{jk}(\vec{k}^{\prime\prime}, \vec{k}_k) \Big[g^{jk}(\vec{k}^\prime_j, \vec{k}_k) \Big]^{-1} \Big[ t^{j} (s_{ru}) \Big]^{-1}.\label{offac}
\end{equation}
\begin{displaymath}
\hspace{7cm} (j\ne r\ne u=1,2,3)
\end{displaymath}
In Eq. (\ref{eq:G}), $\sqrt{s_{lm}}$ is the invariant mass of the $lm$ pair which, as mentioned above, can be calculated in terms of the variables of the formalism. 
The upper index $k^{\prime\prime}$ in the invariant mass $s^{k^{\prime\prime}}_{ru}$ of Eq. (\ref{offac}) indicates a dependence with the loop variable as it was shown in \cite{us_ppn1}.
The definition of 
$G^{i\,j\,k}$ is such that the contribution from diagrams with three $t$-matrices can be written as
\begin{equation}
t^i (s_{lm}) G^{ijk} t^j(s_{ru}) g^{jk}(\vec{k}^\prime_j, \vec{k}_k) t^k(s_{pq})
\end{equation}
which upon substitution of the $G^{i\,j\,k}$ becomes
\begin{eqnarray}\nonumber
&=& t^i (s_{lm}) \biggr( \int\frac{d^3 k^{\prime\prime}}{(2\pi)^3}\tilde{g}^{ij} F^{i\,j \,k} \biggr) t^j(s_{ru}) g^{jk}(\vec{k}^\prime_j, \vec{k}_k) t^k(s_{pq})\quad  (k\ne p\ne q=1,2,3)\nonumber\\
&=&t^i (s_{lm}) \biggr( \int\frac{d^3 k^{\prime\prime}}{(2\pi)^3}\tilde{g}^{ij} t^{j}(s^{k^{\prime\prime}}_{ru}) g^{jk}(\vec{k}^{\prime\prime}, \vec{k}_k) \Big[g^{jk}(\vec{k}^\prime_j, \vec{k}_k) \Big]^{-1}  \\\nonumber
&&\Big[ t^{j} (s_{ru}) \Big]^{-1} \biggr)  t^j(s_{ru}) g^{jk}(\vec{k}^\prime_j, \vec{k}_k) t^k(s_{pq})\\
&=&t^i (s_{lm}) \biggr( \int\frac{d^3 k^{\prime\prime}}{(2\pi)^3}\tilde{g}^{ij} t^{j}(s^{k^{\prime\prime}}_{ru}) g^{jk}(\vec{k}^{\prime\prime}, \vec{k}_k) 
 \biggr) t^k(s_{pq}).
\end{eqnarray}
This shows that in the loops there is a $t$-matrix with an $s^{k^{\prime\prime}}_{ru}$ argument related to the loop variable $k^{\prime\prime}$  and the $k^{\prime\prime}$ integral of all $k^{\prime\prime}$ dependent functions is denoted as $G^{i\,j\,k}$. The next higher order contribution is  written as  $t^i G^{ijk} t^j G^{jkl} t^k g^{kl} t^l$, which was numerically shown to be close to that of an exact calculation in \cite{us_ppn1}. In this way  Eqs. (\ref{Trest}) are algebraic equations. The remaining inputs required to solve Eqs. (\ref{Trest}) are the $J/\psi \pi$, $J/\psi K$, $J/\psi \bar{K}$, $\pi \pi$ and $K \bar{K}$ $t$-matrices, the details of which we discuss in the following subsections.

\subsection{The t-matrix for the pseudoscalar-vector meson interaction.}

For constructing the pseudoscalar-vector interaction Lagrangian we follow the works in \cite{dani1,dani2}. The starting point for the construction of the Lagrangian are fields containing all pseudoscalar and vector mesons from a 15-plet of $SU(4)$ plus a singlet. In the physical basis these fields read:

\be
\Phi&=&\left(
\begin{array}{cccc}
 \frac{\eta }{\sqrt{3}}+\frac{\pi^0}{\sqrt{2}}+\frac{\eta'
   }{\sqrt{6}} & \pi ^+ & K^+ & \overline{D}^0 \\& & & \\
 \pi ^- & \frac{\eta }{\sqrt{3}}-\frac{\pi
   ^0}{\sqrt{2}}+\frac{\eta'}{\sqrt{6}} & K^0 & D^- \\& & & \\
 K^- & \overline{K}^0 & \sqrt{\frac{2}{3}} \eta'-\frac{\eta
   }{\sqrt{3}} &  {D_s}^- \\& & & \\
 D^0 & D^+ &  {D_s}^+ & \eta _c
\end{array}
\right) \\
\cal{V}_\mu&=&\left( \begin{array}{cccc}
{\rho_\mu0 \over \sqd}+{\omega_\mu \over \sqd} & \rho^+_\mu & K^{*+}_\mu & \bar D^{*0}_\mu \\ & & & \\
\rho^{*-}_\mu & {-\rho0_\mu \over \sqd}+{\omega_\mu \over \sqd} & K^{*0}_\mu & D^{*-}_\mu \\& & & \\
K^{*-}_\mu & \bar K^{*0}_\mu & \phi_\mu & D_{s\mu}^{*-} \\& & & \\
D^{*0}_\mu & D^{*+}_\mu & D_{s\mu}^{*+} & J/\psi_\mu \\ \end{array} \right).
\ee
These two fields differ from those used in \cite{dani1,dani2} because of the inclusion of a $SU(4)$ singlet in order to take into account the $\eta$-$\eta'$ and $\omega$-$\phi$ mixing, which was not considered in these previous works.

For each one of these fields a current is defined:

\be
J_\mu&=&(\dmu \Phi)\Phi-\Phi\dmu\Phi \\
\cal{J}_\mu&=&(\dmu \cal{V}_\nu)\cal{V}^\nu-\cal{V}_\nu\dmu \cal{V}^\nu. 
\ee

The Lagrangian is constructed by coupling these currents:

\be
\lc_{PPVV}&=&-{1\over 4f^2}Tr\left(J_\mu\cal{J}^\mu\right). \label{lag}
\ee

The Lagrangian  in Eq. (\ref{lag}) is $SU(4)$ symmetric by construction. We know, though, that $SU(4)$ symmetry is badly broken in nature, because of the heavy charmed quark mass. The first step  to break the $SU(4)$ symmetry in the Lagrangian is to recognize that the interaction behind the coupling of the two currents in Eq. (\ref{lag}) is the exchange of a vector meson, which can be formally visualized within the hidden gauge approach of \cite{hidden1,hidden2,hidden3,ulfvec}. In this way we suppress terms in the Lagrangian where a heavy meson is exchanged by the factor $\gamma=m^2_L/m^2_H$ where $m_L$ is the value of a light vector-meson mass (800 MeV) and $m_H$ the value of the heavy vector-meson mass (2050 MeV).  In the interaction of only heavy mesons ($D^*\bar{D}_{s}$, $\bar{D}_{s} D^*$) the vector meson exchanged is the $J/\psi$ and we suppress it by the factor $\psi=m^2_L/m^2_{J/\psi}$. We also consider different values for the $f$ appearing in the coupling of Eq. (\ref{lag}), for light mesons we use $f=f_\pi=93$ MeV but for heavy ones $f=f_D=165$ MeV.

With our phenomenological Lagrangian we can obtain the potential for a given process $(P(p)V(k))_i\rightarrow (P'(p')V'(k'))_j$:

\be
v_{ij}(s,t,u)&=&-{\xi_{ij}\over4f_i f_j}(s-u)\epsilon . \epsilon ', \label{ampli}
\ee
where $s$ and $u$ are the usual Mandelstam variables, $f_k$ is the decay constant of the  pseudoscalar  meson $k$, $\epsilon$'s are the vector-meson polarization vectors and $i$, $j$ refer to the initial and final channels in the coupled channel space. The matrix of coefficients  $\xi_{ij}$ can be directly calculated from the Lagrangian of Eq. (\ref{lag}). Eq. (\ref{ampli}) should be projected into s-wave, which is the only partial wave that we study. We come back to technical details in the results section.

We take the following coupled channels for the strangeness S=1 case:
 $K^{*}\pi$, $\rho K$, $K^{*}\eta$, $K^{*}\eta\prime$, $\omega K$,
$\phi K$, $D_s^{*}\bar D$, $\bar D^{*}D_s$, $J/\psi K$ and $K^{*}\eta_c$. And the coefficient matrix $\xi_{ij}$ for these channels in isospin $I=\frac{1}{2}$ is given below

\be
\xi&=&
\left(
\begin{array}{llllllllll}
 2 & -\frac{1}{2} & 0 & 0 & -\frac{\sqrt{3}}{2} & \sqrt{\frac{3}{2}} &
\sqrt{\frac{3}{2}} \gamma  & 0 & 0 & 0 \\
 -\frac{1}{2} & 2 & \sqrt{2} & -\frac{1}{2} & 0 & 0 & 0 & -\sqrt{\frac{3}{2}} \gamma
 & 0 & 0 \\
 0 & \sqrt{2} & 0 & 0 & -\sqrt{\frac{2}{3}} & \frac{2}{\sqrt{3}} & \frac{\gamma
}{\sqrt{3}} & -\frac{\gamma }{\sqrt{3}} & 0 & 0 \\
 0 & -\frac{1}{2} & 0 & 0 & \frac{1}{2 \sqrt{3}} & -\frac{1}{\sqrt{6}} &
\frac{\gamma }{\sqrt{6}} & \sqrt{\frac{2}{3}} \gamma  & 0 & 0 \\
 -\frac{\sqrt{3}}{2} & 0 & -\sqrt{\frac{2}{3}} & \frac{1}{2 \sqrt{3}} & 0 & 0 & 0 &
\frac{\gamma }{\sqrt{2}} & 0 & 0 \\
 \sqrt{\frac{3}{2}} & 0 & \frac{2}{\sqrt{3}} & -\frac{1}{\sqrt{6}} & 0 & 0 & \gamma 
& 0 & 0 & 0 \\
 \sqrt{\frac{3}{2}} \gamma  & 0 & \frac{\gamma }{\sqrt{3}} & \frac{\gamma
}{\sqrt{6}} & 0 & \gamma  & \psi  & 0 & -\gamma  & -\gamma  \\
 0 & -\sqrt{\frac{3}{2}} \gamma  & -\frac{\gamma }{\sqrt{3}} & \sqrt{\frac{2}{3}}
\gamma  & \frac{\gamma }{\sqrt{2}} & 0 & 0 & \psi  & -\gamma 
   & -\gamma  \\
 0 & 0 & 0 & 0 & 0 & 0 & -\gamma  & -\gamma  & 0 & 0 \\
 0 & 0 & 0 & 0 & 0 & 0 & -\gamma  & -\gamma  & 0 & 0
\end{array}
\right) \nn
\ee

For strangeness S=-1 the coupled channels considered are 
$\bar{K^{*}}\pi$, $\rho \bar{K}$, $\bar{K^{*}}\eta$, $\bar{K^{*}}\eta\prime$, $\omega \bar{K}$,
$\phi \bar{K}$, $\bar{D_s^{*}} D$, $D^{*}\bar{D_s}$, $J/\psi \bar{K}$ and $\bar{K^{*}}\eta_c$
and the coefficients for these channels are the same as for their corresponding S=1 channels above.

For strangeness S=0, one can find the coupled channels and the coefficient matrix in \cite{meux3872}.

To obtain the $t$-matrix we project in s-wave the potentials of Eq. (\ref{ampli}) (removing -$\epsilon\cdot\epsilon^\prime$) and plug them into the scattering equation for the coupled channels:

\be
t&=& v + v g^\prime t. \label{bseq}
\ee
In this equation $g^\prime$ is a diagonal matrix with each one of its elements given by the loop function for each channel in the coupled channel space. For channel $i$ with mesons of masses $m_1$ and $m_2$, $g^\prime_{ii}$ is given by:

\be
g^\prime_{ii}&=&{1 \over 16\pi ^2}\biggr( \alpha_i + Log{m_{1}^{2} \over \mu ^2}+{m_{2}^{2}-m_{1}^{2}+s\over 2s}
  Log{m_{2}^{2} \over m_{1}^{2}}\nonumber\\ 
 &+ &{p\over \sqrt{s}}\Big( Log{s-m_{2}^{2} + m_{1}^{2} + 2p\sqrt{s} \over -s + m_{2}^{2} - m_{1}^{2} +
  2p\sqrt{s}}\nn\\
&+&Log{s+m_{2}^{2} - m_{1}^{2} + 2p\sqrt{s} \over -s - m_{2}^{2} + m_{1}^{2} + 2p\sqrt{s}}\Big)\biggr),
\ee
where $p$ is the three momentum of the two mesons in their center of mass frame. The two parameters $\mu$ and $\alpha$ are not independent, we fix $\mu$=1500 MeV and use for $\alpha$ the same values used in \cite{dani2}. These values of $\alpha$ are obtained from moderate changes from their natural size \cite{ollerulf} in order to fit the spectrum for most of the known light and charmed axial resonances.

\subsection{The t-matrix for the pseudoscalar-pseudoscalar meson interaction.}
The $\pi \pi$, $K \bar{K}$ diagonal and non-diagonal potential has been obtained from the lowest order chiral Lagrangian \cite{oller}
\begin{equation}
\lc = \frac{1}{12f^2} Tr\big( (\dmu \Phi\Phi -\Phi\dmu\Phi)^2 + M\Phi^4 \big), \label{lagpp}
\end{equation}
where,
\be
\Phi&=&\left(
\begin{array}{ccc}
\frac{\pi0}{\sqrt{2}}+ \frac{\eta_8}{\sqrt{6}}& \pi ^+ & K^+  \\
& &  \\
 \pi ^- & -\frac{\pi^0}{\sqrt{2}}+\frac{\eta_8 }{\sqrt{6}} & K^0 
 \\& &  \\
 K^- & \overline{K}^0 & -\frac{2}{\sqrt{6}} \eta_8 
\end{array}
\right)
\ee
and
\be
M&=&\left(
\begin{array}{ccc}
m_\pi^2 & 0 & 0  \\
& &  \\
 0 & m_\pi^2  & 0 
 \\& &  \\
 0 & 0 & 2 m_K^2 - m_\pi^2
\end{array}
\right).
\ee
The on-shell part of the potential obtained from the Lagrangian Eq. (\ref{lagpp}) in $S$-wave, for total isospin  of the two pseudoscalars equal to 0, is \cite{oller} 
\begin{eqnarray}\nonumber
V_{K\bar{K} \rightarrow K\bar{K}} &=& - \frac{3}{4 f^2} \,\, s_{23}\\\nonumber
V_{\pi\pi \rightarrow K\bar{K}} &=& - \frac{\sqrt{3}}{4 f^2} s_{23}\\\label{potpp}
V_{\pi\pi \rightarrow \pi\pi} &=& - \frac{1}{f^2} \Biggr( s_{23} - \frac{m_\pi^2}{2} \Biggr).
\end{eqnarray}
These potentials are used to solve the Bethe-Salpeter equations for $\pi\pi$ and $K\bar{K}$ coupled channels using the same subtraction constants as the ones used in \cite{oller}. We would like to mention that we have taken care of the symmetrization of the $\pi \pi$ states. The dynamical generation of the $\sigma$ and $f_0 (980)$ scalar resonances in these systems was found using the potentials Eq. (\ref{potpp}) in \cite{oller}. 

Thus, we obtain the $t$-matrices for the scattering of two pseudoscalars and of the vector-pseudoscalar mesons which reproduce the experimental data in the corresponding cases. With these inputs we solve the Faddeev equations Eqs. (\ref{Trest}). We shall now discuss the results of our calculations.

\section{Results and conclusions}
\begin{figure}[h!]
\begin{center}
\includegraphics[width=8cm]{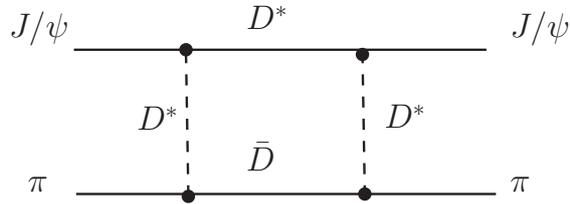}
\caption{The $J/\psi \pi\to J/\psi\pi  $ amplitude proceeding through the intermediate $D^* \bar{D}$ channel in the loop with a $D^*$ exchange at the $J/\psi \pi \to D^* \bar{D}$ vertex.}
\label{dstarexch}
\end{center}
\end{figure}

Using the $t$-matrices explained in the above section as input, we solve Eqs. (\ref{Trest}) for the $J/\psi \pi \pi$ and  $J/\psi K \bar{K}$ channels in total isospin 0, varying  the total energy $\sqrt{s}$ between 4 and 5 GeV and the invariant mass of the two pseudoscalars, $\sqrt{s_{23}}$,  between 400 to 1100 MeV. 
As explained above, the $J/\psi \pi$ and $J/\psi K$ interaction is null at the lowest order but it is non-zero when the loops of the coupled channels are considered in the iteration of the potential leading to the $t$-matrix. A diagram for the lowest order non-zero contribution to the $J/\psi \pi$ interaction has been shown in Fig. \ref{fig1}, and its contribution is written mathematically as
\begin{equation}
v_{\pi\pi\to\pi\pi} g^{13}\, v_{J/\psi \pi \rightarrow D^* \bar{D}} \,  g^\prime_{D^* \bar{D}} \, v_{D^* \bar{D} \rightarrow J/\psi \pi}.  \label{vdd}
\end{equation}
The potential in Eq. (\ref{ampli}) has been obtained by assuming that the momentum transfer, i.e., the Mandelstam variable $t$, in $J/\psi \pi \rightarrow D^* \bar{D}$ amplitude is negligibly small compared to the vector mass. However for the energies and channels considered here, such an approximation is not good and we need to take the effect of large momentum transfer into account. In order to do this, we consider the $D^*$-exchange in the  $J/\psi \pi \rightarrow D^* \bar{D}$ potential (following \cite{inoue}) to get
\begin{equation}
v_{J/\psi \pi \rightarrow D^* \bar{D}} \to \int \frac
{\hat{dk^\prime}}{4\pi}v_{J/\psi \pi \rightarrow D^* \bar{D}}  \frac{-m_{D^*}^2}{(k^\prime- k)^2 - m_{D^*}^2},
\end{equation}
where $k^\prime$ and $k$ are the four vectors of the $D^*$ and the $J/\psi$, respectively.
This would mean that the $J/\psi \pi\to J/\psi \pi$ amplitude implicit in Eq. (\ref{vdd}) would be as shown in Fig.\ref{dstarexch}. Similarly, we take into account this correction for the $J/\psi K$ and the $J/\psi \bar{K}$
amplitudes also.

With this new potentials we calculate the $t$-matrix for the $J/\psi$-pseudoscalar interaction and carry out the calculations for the the $J/\psi \pi \pi$ and the $J/\psi K \bar{K}$ systems. We find a resonance in both the systems at $\sqrt{s}$ = 4150 MeV with a full width at half maximum of 90 MeV. The peak appears when the invariant mass of two pseudoscalars is around that of the $f_0 (980)$, indicating that the resonance has a strong coupling to the $J/\psi f_0 (980)$ channel. Both the $J/\psi \pi \pi$ and the $J/\psi K \bar{K}$ amplitudes are similar in this energy region, with a difference in their magnitudes. We find the $J/\psi K \bar{K}$ amplitude to be much larger in magnitude as compared to that of the $J/\psi \pi \pi$ system. This reveals the strong coupling of the three-body resonance to $J/\psi f_{0}(980)$, since the $f_{0}(980)$ couples most strongly to $K\bar{K}$ \cite{pelaez,ablikim2,kaiser}.  

\begin{figure}[h!]
\begin{center}
\includegraphics[width=12cm]{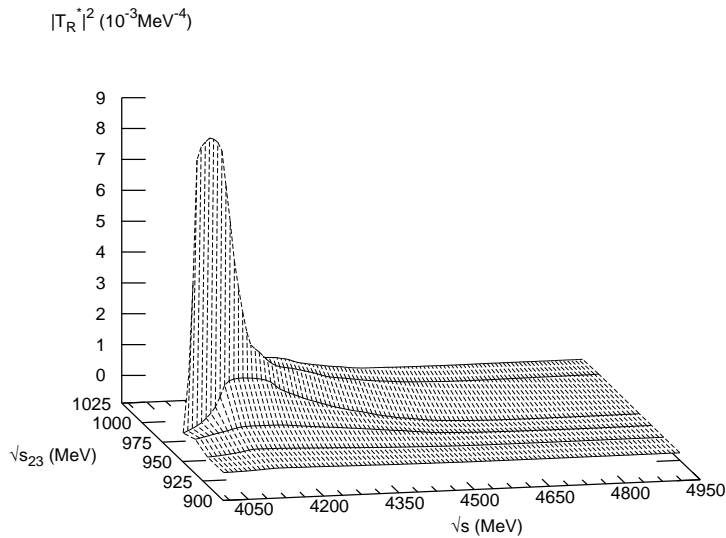}
\caption{$|T^*_{R}|^2$ for the $J/\psi K\bar{K}$ system in total isospin $I=0$ as a function of the total energy, $\sqrt{s}$, and the invariant mass of the $K\bar{K}$ subsystem, $\sqrt{s_{23}}$.}
\label{4260}
\end{center}
\end{figure}

In Fig. \ref{4260} we show the $J/\psi K \bar{K}$ squared amplitude as a function of the total energy of the three body system and the invariant mass of the $K \bar{K}$ system.
We have also studied the invariant mass spectrum of the two pions at $\sqrt{s}$ = 4 GeV, 4.3 GeV and 4.5 GeV, i.e., in the energy region of the resonance and below and above it. To do that we take the three-body $|T^*_{R}|^2$-matrix and multiply it by the phase space factor 
\begin{equation}
\frac{\tilde{p}\cdot \tilde{q}}{\sqrt{s}}
\end{equation}
where $\tilde{p}=\frac{\lambda^{1/2}(s,m^2_{J/\psi},M^2_{inv})}{2\sqrt{s}}$ is the momentum of the $J/\psi$ in the global center of mass system and $\tilde{q}=\frac{\lambda^{1/2}(M^2_{inv},m^2_{\pi},m^2_{\pi})}{2M_{inv}}$ the momentum of the pion in the corresponding two-body center of mass system ($M_{inv}$ is  the invariant mass of the two pions).

 As shown in Figs. \ref{inv2}, the invariant mass spectrum at  $\sqrt{s}$ = 4 GeV shows a phase space like behavior and the one at  $\sqrt{s}$ = 4.3 GeV shows a dominant peak of the $f_0 (980)$ resonance. At 4.5 GeV, we still see the presence of the $f_0 (980)$ in the two pion mass spectrum but  the  magnitude of this peak is much smaller as compared to the one seen at $\sqrt{s}$ = 4.3 GeV, and we find that it gradually  fades away at higher energies.

\begin{figure}
\vspace{-2cm}
\hspace{-2.cm}
\includegraphics[scale=0.35]{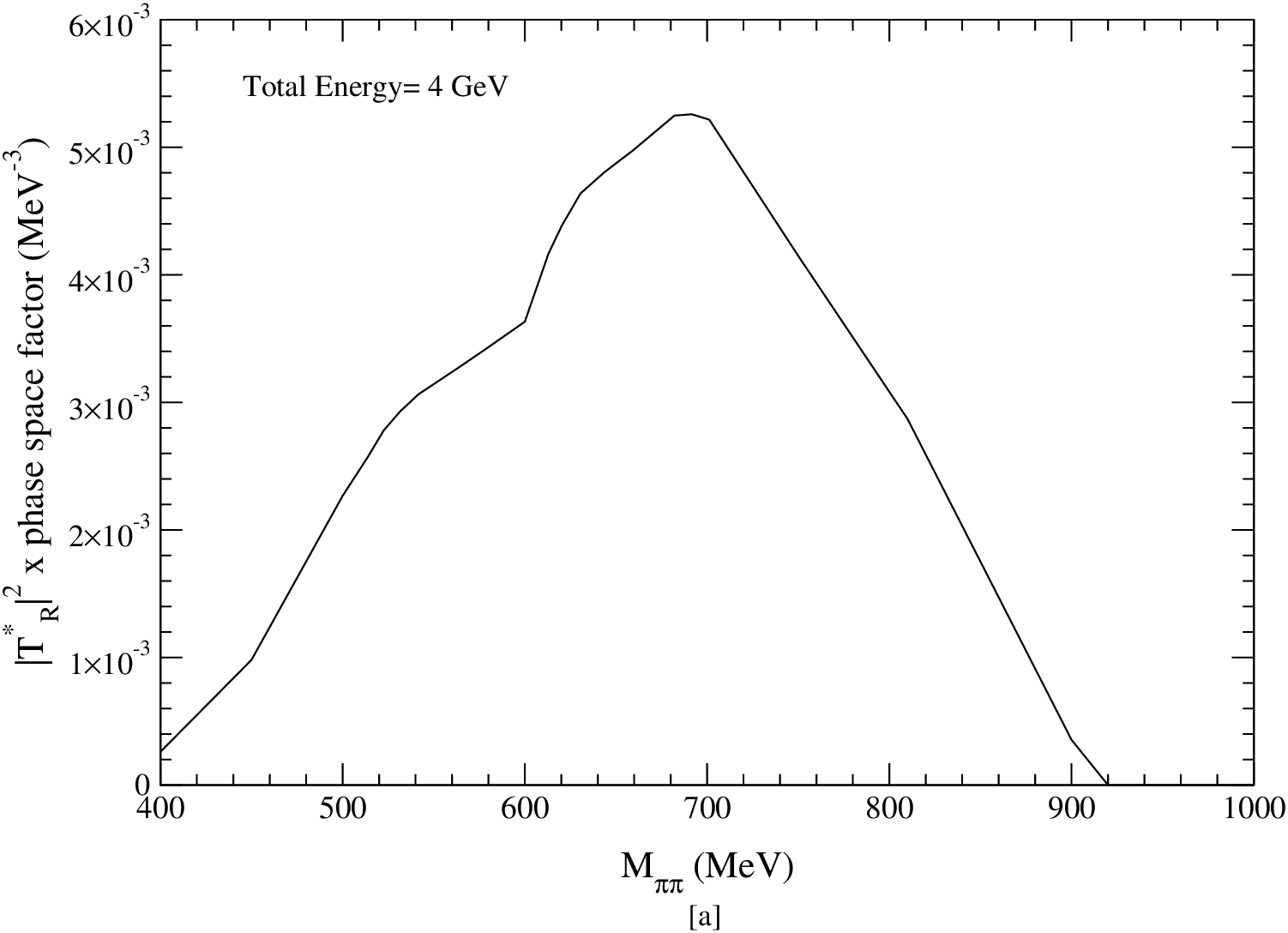}
\quad
\includegraphics[scale=0.35]{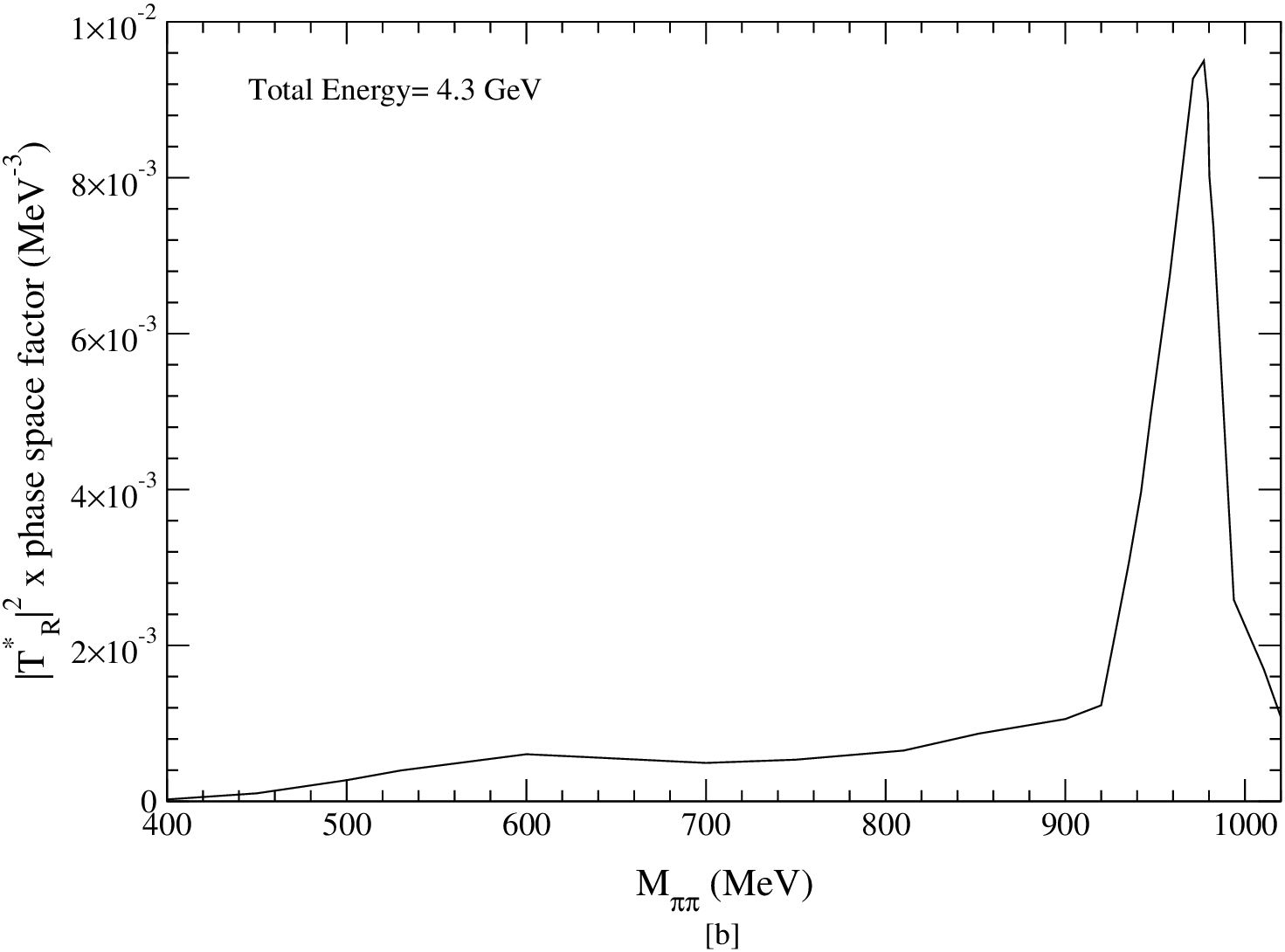}
\vspace{2.cm}
\end{figure}
\begin{figure}
\vspace{-2.36cm}
\centering
\includegraphics[scale=0.35]{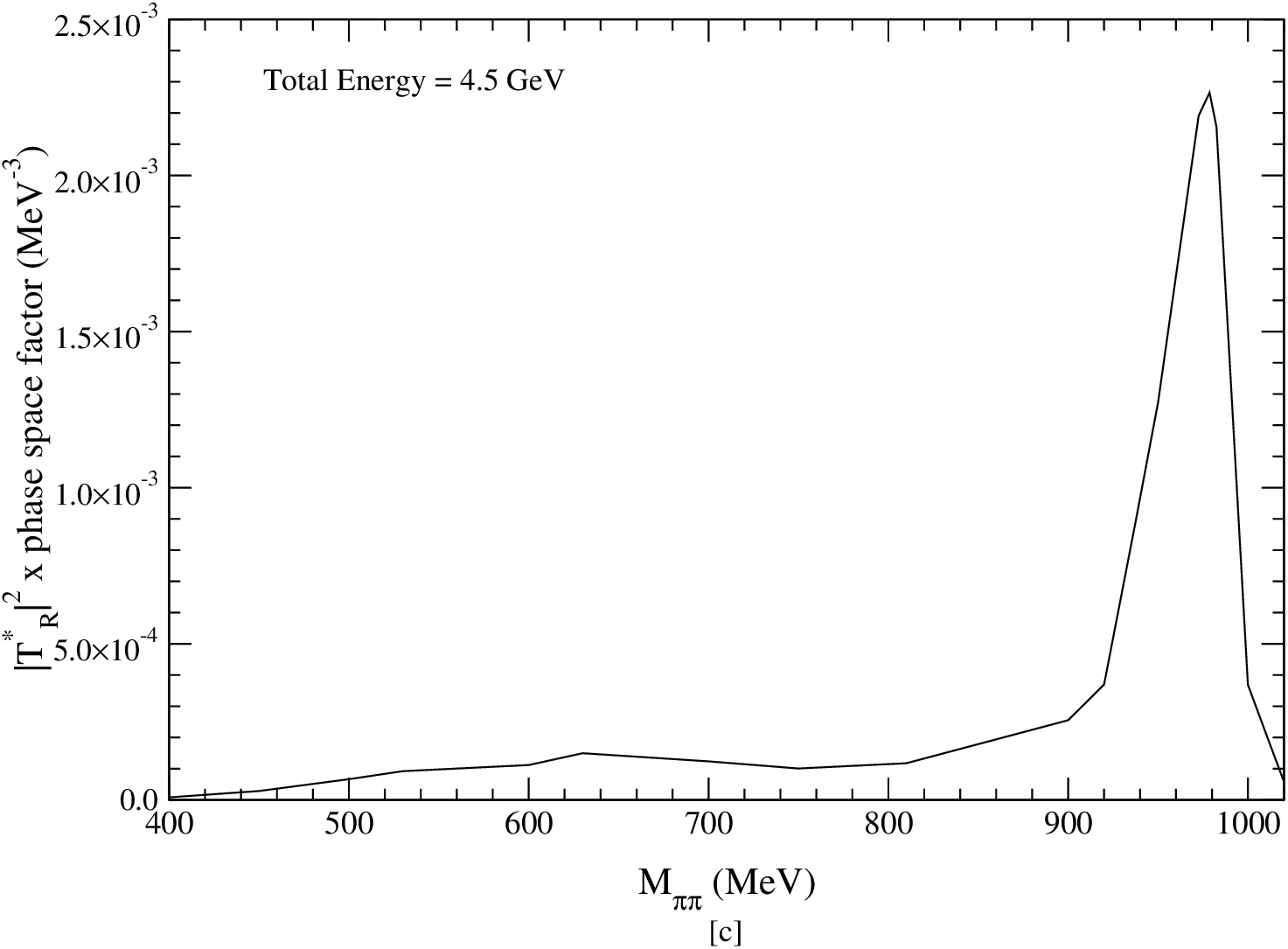}
\caption{$|T^*_{R}|^2$ times the phase space factor for $J/\psi \pi\pi$ plotted as a function of the invariant mass $M_{\pi\pi}$ of the two pions system for three different total energies: a) 4 GeV; b) 4.3 GeV; c) 4.5 GeV.}\label{inv2}
\end{figure}

The features described above and depicted in Fig. \ref{inv2} agree qualitatively with those found for the $M_{\pi\pi}$ spectrum in \cite{belle1}. One should note that the peak of the $|T|^2$ matrix is found around $4150$ MeV rather than the nominal $4260$ MeV. While $100$ MeV difference is not a big difference for a hadronic model where no parameters have been fitted to the resonance data, the fact remains that this difference is the largest one found so far for all the three-body states that we have studied before \cite{us2175,us_pkbarn1,us_ppn1,us_ppn2,us_pkn}. This should be not surprising and we would like to attribute it to uncertainties in $SU(4)$ and the fact that, unlike other cases, here we have no data to tune our $J/\psi \pi$ and $J/\psi K$ ($\bar{K}$) interaction with our limited freedom in the subtraction constants.

In order to have some rough estimate of uncertainties we have varied the SU(4) symmetry breaking parameter, $\gamma$, which enters the evaluation of the $J/\psi \pi\to J/\psi \pi$ or $J/\psi K(\bar{K})\to J/\psi K (\bar{K})$ amplitudes, which proceed as shown in Fig. \ref{dstarexch} and involve necessarily this parameter. We summarize the results here: if we increase $\gamma$ in 15 $\%$ we find that the strength of the peak of Fig. \ref{4260} is also increased in about 50 $\%$. The magnitud of the peaks in Fig. \ref{inv2} are also changed in a similar amount. However, we see that the position of the peaks and their widths are affected much less and the changes found are of the order of 5 MeV for both.

To summarize the results, the quantum numbers of the state obtained, the proximity in the mass to the experimental one and particularly the decay mode of the resonance give us strong reasons to associate the state found to the $Y(4260)$ resonance.

\section*{Acknowledgments}  
This work is partly supported by DGICYT contract number
FIS2006-03438 and  the JSPS-CSIC
collaboration agreement no. 2005JP0002, and Grant for Scientific
Research of JSPS No.188661.
One of the authors (A. M. T) is supported by a FPU grant of the Ministerio de Ciencia e Innovaci\'on.
K. P. Khemchandani thanks the support by the
\textit{Funda\c{c}$\tilde{a}$o para a Ci$\hat{e}$ncia e a Tecnologia of the Minist\'erio da Ci$\hat{e}$ncia, Tecnologia e Ensino Superior} of Portugal (SFRH/BPD/40309/2007). This research is  part of
the EU Integrated Infrastructure Initiative  Hadron Physics Project
under  contract number RII3-CT-2004-506078.

\end{document}